\documentclass[%
 reprint,
 amsmath,amssymb,
 aps,
 nofootinbib
]{revtex4-2}

\usepackage{graphicx}
\usepackage{dcolumn}
\usepackage{bm}

\usepackage{appendix}
\usepackage[unicode=true,
 bookmarks=true,bookmarksnumbered=false,bookmarksopen=false,
 breaklinks=false,pdfborder={0 0 1},backref=false,colorlinks=true]
 {hyperref}
\usepackage{breakurl}
\usepackage[hang,flushmargin]{footmisc}
\usepackage{color}
\usepackage{graphicx}

\hypersetup{pdftitle={Predictions for Quantum Gravitational Signatures from Inflation},
 pdfauthor={Aidan Chatwin-Davies, Achim Kempf, Petar Simidzija},
 citecolor=black,linkcolor=black,urlcolor=black}
\newcommand{\altemail}[1]{\href{mailto:#1}{\nolinkurl{#1}}}

\newcommand{\Eq}[1]{Eq.~\eqref{#1}}

\newcommand{\mrm}[1]{\mathrm{#1}}
\newcommand{\dee}{\mathrm{d}}

\newcommand{\calR}{\mathcal R}

\newcommand{\DR}{\Delta_{\mathcal R}^2}
\newcommand{\dDR}{\delta\Delta_{\mathcal R}^2}

\newcommand{\Mpl}{M_\mrm{Pl}}
\usepackage{xcolor}
\usepackage[normalem]{ulem}
\definecolor{purple}{rgb}{0.5,0,0.5}

\begin{document}

\title{Predictions for Quantum Gravitational Signatures from Inflation}

\author{Aidan Chatwin-Davies${}^{a,b}$}
\author{Achim Kempf${}^{c}$}
\author{Petar Simidzija${}^{a}$}%
 \altaffiliation{\altemail{achatwin@phas.ubc.ca}\\ \altemail{akempf@pitp.ca}\\\altemail{psimidzija@phas.ubc.ca}\\~}
\affiliation{~\\${}^a$Department of Physics and Astronomy, University of British Columbia\\
 Vancouver, BC, V6T 1Z1, Canada\\~\\
 ${}^b$Institute for Theoretical Physics, KU Leuven\\
 Celestijnenlaan 200D B-3001 Leuven, Belgium \\~\\
 ${}^c$Department of Applied Mathematics, University of Waterloo\\
 Waterloo, ON, N2L 3G1, Canada\\~
}

\date{\today}

\begin{abstract}
We compute the corrections to the primordial power spectrum that should arise in realistic inflationary scenarios if there exists a generic covariant ultraviolet (UV) cutoff, as commonly motivated by considerations of quantum gravity.
The corrections to the spectrum consist of small superimposed oscillations whose frequency, phase, and amplitude are functions of the comoving wave number. For any given cosmological parameters that characterize the slow roll during inflation, the frequency predicted for these oscillations depends only on the value of the UV cutoff.
The specificity of this prediction can be used to increase experimental sensitivity through the filtering for template signatures. This will allow experiments to put new bounds on where a natural UV cutoff can be located between the Planck scale and the Hubble scale during inflation. It may even bring imprints of Planck-scale physics in the cosmic microwave background and in structure formation within the range of observations.

~

\end{abstract}

\maketitle

\section{Introduction}

It is widely expected that a theory of quantum gravity will be needed to describe physics  at the Planck scale. 
There are several competing candidates for such a theory, see, e.g., \cite{Rovelli:1997qj,WittenStrings,Carlip:2015asa,Loll:2022ibq}, but experiments that could probe Planck-scale physics  are still lacking. For example, particle accelerators are presently able to operate at centre-of-mass energies that are still about 15 orders of magnitude below the Planck scale.

Fortunately, observational early universe cosmology involves scales that are substantially closer to the Planck scale. This is because, according to standard inflationary theory, the fluctuations that are now visible in the cosmic microwave background (CMB) originated in primordial quantum fluctuations that occurred when the Hubble length during inflation was only 5 to 6 orders of magnitude away from the Planck length.\footnote{See Eq.~\eqref{eq:Heff}.} This raises the exciting question of what type of observational signature Planck-scale physics may have left in the CMB, and at what magnitude.

Even without a confirmed UV-complete theory of quantum gravity at hand, it is clear that, as the Planck scale is approached from lower energies, conventional quantum field theory should start to exhibit modifications that are due to Planck-scale physics \cite{Garay:1994en,Hossenfelder:2012jw}. To some extent, these modifications should have  impacted the inflationary mechanism for the generation of primordial fluctuations. 
In the literature, the possibility of Planck-scale signatures in the CMB has been addressed, therefore, by modelling natural ultraviolet (UV) cutoffs, associated modifications to dispersion relations, and other approaches \cite{Padmanabhan:1988jp,Padmanabhan:1988se,Jacobson:1999zk,Kempf:2000ac,Martin:2000xs,Brandenberger:2000wr,Niemeyer:2000eh,Brandenberger:2001zqv,Easther:2001fi,Kempf:2001fa,Easther:2001fz,Brandenberger:2002hs,Easther:2002xe,Danielsson:2002kx,Brandenberger:2004kx,Sriramkumar:2004pj,Greene:2004np,Shiu_2005,Easther:2005yr,Tawfik:2015rva,Ali:2015ola,Skara:2019uzz,Frob:2012ui,Frob:2014cza,Calcagni:2016ofu,Calcagni:2017via,Modesto:2022asj,Calcagni:2022tuz,Kaloper:2002uj,deBlas:2016puz,Ashtekar:2020gec,Agullo:2021oqk,Ashtekar:2021izi,Navascues:2021mxq,Navascues:2021qcp} within the quantum field theoretic framework of inflation. 
However, most of these approaches break local Lorentz invariance, which makes it difficult to distinguish to what extent the predicted effects would be due to Planck-scale physics and to what extent they would be due to the symmetry breaking. 

In previous work, we therefore introduced into inflation a class of natural UV cutoffs that are fully covariant. The requirement of covariance is technically hard to implement but has the benefit of being highly restrictive. We then showed how the conventional inflationary calculations, when augmented with a natural covariant cutoff, could be used, in principle, to compute a prediction for Planckian signatures in cosmological observables \cite{Chatwin-Davies:2016byj}.

In the present letter, and in a longer article \cite{Chatwin-Davies:2022irs}, we build on this prior work and 
compute the effect of covariant natural UV cutoffs on the primordial power spectra (PPS) of scalar and tensor fluctuations for realistic single-field inflation models. 

We find that the presence of a covariant UV cutoff (either soft or sharp) results in small oscillations superposed on the conventional PPS, and we arrive at an explicit prediction for the frequency of these oscillations.
Their frequency is a function, Eq.~\eqref{eq:frequency}, of a single new dimensionless parameter: the ratio $\sigma$ of the Hubble parameter at horizon crossing, $H$, and the cutoff scale, $\Omega$.

In the simplest case, where the natural UV cutoff is a sharp cutoff, we can predict not only the frequency, but also the amplitude and phase of the oscillatory perturbations. This yields a family of possible template signatures in the CMB that is parametrized only by the value of the natural UV cutoff scale in addition to standard cosmological parameters. The high specificity of the predicted signatures suggests that experimental sensitivity can be significantly enhanced by 
using template matching techniques, similarly to how current gravitational wave detectors are able to detect the presence of extremely low amplitude gravitational waves. Using the new results and comparing them with PPS data, it should at least be possible to put new stringent bounds on where, in between the Hubble length during inflation and the Planck length, a natural UV cutoff can be located.

\section{Primordial power spectra}

In single-field inflationary models, the rapid expansion is driven by a single scalar field, the \textit{inflaton}, rolling down its potential. After taking into account that the quantum fluctuations of the inflaton couple to the quantum fluctuations of the effectively scalar part of the metric, one obtains three decoupled degrees of freedom: a free scalar $\calR$ and two free tensor helicities $\cal T_\pm$. The quantum fluctuations of $\calR$ and $\cal T_\pm$ are thought to be the origin of all inhomogeneities in the universe. In this letter, we will mainly focus on the scalar $\calR$. The tensor perturbations, which have not been observed as yet, can be studied similarly; for details, see \cite{Chatwin-Davies:2022irs}.

It is convenient to decompose the free field $\calR$ into its comoving Fourier modes. For a mode with a wavevector of magnitude $k$, the variance of its quantum fluctuations is quantified by the PPS $\Delta_{\calR}(k)$ \cite{dodelson2003modern}. Measurements of the CMB find that the PPS is of the form
\begin{align} \label{eq:scalar-spectrum-pheno}
    \Delta_{\mathcal R}^2(k) = A_s\left(\frac{k}{k_\star}\right)^{n_s-1},
\end{align}
where, at the pivot scale $k_\star = 0.05~\mrm{Mpc}^{-1}$, the spectral amplitude and spectral tilt are $A_s = (2.10\pm0.03)\times 10^{-9}$ and $n_s = 0.966 \pm 0.004$, respectively \cite{Planck:2018vyg}. Because the spectral tilt is close to $1$, the PPS is nearly scale-invariant. 

Such a scale-invariant PPS can be obtained if the inflaton field which drives inflation is {slowly rolling} down its potential, i.e. if the slow roll parameter $\epsilon \equiv -\dot H/H^2$ is much less than 1, where $H$ is the Hubble parameter and a dot denotes a derivative with respect to cosmic time $t$. In this case, the spacetime is quasi-de Sitter and the PPS evaluates to (see, e.g., \cite{dodelson2003modern})
\begin{equation} \label{eq:scalar-spectrum-slow-roll}
    \Delta_{\cal R}^2(k) = \frac{H^2(\eta_k)}{\pi M_P^2 \epsilon{(\eta_k)}},
\end{equation}
where $\eta_k$ is the comoving time at which a mode $k$ crosses the Hubble length.

\section{Covariant ultraviolet cutoffs}

The primordial power spectrum $\DR(k)$, being a measure of two-point correlations of the field $\calR$, can also be written in terms of the spatial Fourier transform $G_F(\eta,k)$ of the Feynman propagator for $\calR$:
\begin{equation} \label{eq:PSP-GF}
    \Delta_{\calR}^2(k) = 4\pi k^3 |G_F(\eta_k,k)|.
\end{equation}
Writing $\DR(k)$ in this form is useful because the Feynman propagator can be written as a path integral, and the path integral can be naturally modified to include a covariant UV cutoff.

To see this, we start with a path integral as a sum over field configurations $\phi(x)$.
Each field configuration is weighted by the complex phase $\exp(iS[\phi])$, where $S[\phi]$ is the action functional. For example, the Feynman propagator of a real scalar field $\phi$ is then
\begin{equation} \label{eq:GF}
    i G_F(x,x') = \frac{\int \mathcal{D}\phi ~ \phi(x) \phi(x') e^{iS[\phi]}}{\int \mathcal{D}\phi ~ e^{iS[\phi]}}.
\end{equation}
By the stationary phase approximation, the classical field configurations, i.e., the stationary points of the action functional, tend to dominate the path integral.
Meanwhile, the non-stationary contributions represent quantum fluctuations and they tend to be suppressed by averaging out.

This can be made more precise in the case of free field theories, such as those describing inflationary scalar and tensor perturbations. For a massless free field $\phi$ on a manifold $\mathcal M$, the stationarity of the action is equivalent to the linear field equation $\Box \phi = 0$, where $\Box$ is the self-adjoint d'Alembertian on $\mathcal M$. In this case, classical field configurations are those for which $\Box \phi = 0$, while field configurations that describe quantum fluctuations are linear combinations of eigenfunctions of $\Box$ which possess non-zero eigenvalues. 

The magnitude of these eigenvalues determines the extent to which these quantum fluctuations are off shell.
Since the conventional path integral does not bound these eigenvalues, it includes quantum fluctuations that are arbitrarily far off shell, i.e., for which $\Box \phi = \lambda \phi$ where $|\lambda|$ arbitrarily far exceeds the Planck scale.
We therefore obtain a covariant UV cutoff by path integrating only over field configurations obeying $|\lambda|<\Omega^2$, where $\Omega$ sets the UV cutoff scale, while omitting the more extreme, trans-Planckian configurations. 
One may expect $\Omega$ to be at or near the Planck scale, but its value must of course be determined experimentally. 
For a sharp UV cutoff, the set of allowed field configurations in the path integral is then
\begin{equation}
    B_{\mathcal{M}}(\Omega) \equiv \mrm{span}\left\{ \psi_\lambda \, | \, \Box \psi_\lambda = \lambda \psi_\lambda, |\lambda| \leq \Omega^2  \right\}.
\end{equation}
The allowed eigenvalues are bounded from above and below since the spectrum of $\Box$ is unbounded in both the positive and negative directions. Hence, for instance, the cut-off Feynman propagator is given by
\begin{equation} \label{eq:GF-cutoff}
    i G_F^\Omega(x,x') \equiv \frac{\int_{B_\mathcal{M}(\Omega)} \mathcal{D}\phi ~ \phi(x) \phi(x') e^{iS[\phi]}}{\int_{B_\mathcal{M}(\Omega)} \mathcal{D}\phi ~ e^{iS[\phi]}}.
\end{equation}
Note that this spectral cutoff $\Omega$ is manifestly covariant since $\Box$ is a scalar operator and, therefore, the spectrum of $\Box$ is independent of the choice of coordinates for $\mathcal{M}$. 

In the simple example where the manifold $\mathcal{M} = \mathbb{R}$, the scalar wave operator is just $-\partial_x^2$, its eigenfunctions are plane waves $e^{ikx}$, its eigenvalues are the frequencies $k^2$, and hence $\Omega$ is a maximum frequency cutoff. Thus, for general manifolds, we can think of $\Omega$ as a covariant generalization of a maximum frequency.

We remark that the cutoff $\Omega$ also possesses an information theoretic interpretation as a cutoff on the density of field degrees of freedom in spacetime, in the sense of Nyquist-Shannon sampling theory \cite{shannon1998mathematical,NyquistReprint}.
For example, consider again $\mathcal{M} = \mathbb{R}$ and suppose that a function $\phi(x)$ has a Fourier transform that vanishes outside of a window $[-\Omega,\Omega]$.
Then, theorems of Shannon and Beurling state that $\phi(x)$ can be fully reconstructed knowing only the values it takes at a discrete lattice of sample points, whose average density is greater than $\Omega/\pi$ \cite{ShannonTheorem,LandauSampling}.
Similarly, one can also think of a covariant cutoff $\Omega$ as a covariant bandlimit, which controls the density of information contained in $\phi(x)$.
This information theoretic interpretation generalizes to Lorentzian manifolds, where the density of information in space and time transforms covariantly when going from one reference frame to another \cite{Kempf:1999xt,Kempf:2003qu,Kempf:2009us,Kempf:2010rx,Kempf:2012sg}.

Consider again the computation of the covariantly cut-off Feynman propagator $G_F^\Omega$ in Eq. \eqref{eq:GF-cutoff}.
Instead of directly evaluating the path integral over the set of bandlimited fields configurations $B_{\mathcal M}(\Omega)$, it is equivalent and simpler in practice to project out the high eigenvalue contributions to $G_F$ \cite{Chatwin-Davies:2016byj,Chatwin-Davies:2022irs}.
Concretely, we act on $G_F$ to the left and right with projectors, $P_\Omega$: 
\begin{align}\label{eq:G_cutoff}
    G_F^\Omega = P_\Omega G_F P_\Omega.
\end{align}
The projector $P_\Omega$ is a linear operator that acts on fields on $\mathcal{M}$ and that is defined as
\begin{equation} \label{eq:cutoff}
    P_\Omega \equiv \sum_{\lambda \in \mrm{spec}\,\Box} \theta(\Omega^2 - |\lambda|) \langle \psi_\lambda, \, \cdot \, \rangle \psi_\lambda,
\end{equation}
where $\langle \, \cdot \, , \cdot \, \rangle$ denotes the inner product on $\mathcal{M}$ and $\psi_\lambda$ is the eigenfunction of $\Box$ with eigenvalue $\lambda$.
As a result, eigenvalues $|\lambda| > \Omega^2$ are projected out.

This way of expressing the cutoff further illustrates that the sharp cutoff described so far is part of a larger class of covariant cutoffs obtained by replacing the Heaviside step function $\theta$ in \Eq{eq:cutoff} by a general non-negative function $f$.\footnote{A systematic way to study possible high-energy corrections to low-energy observables of a field theory is via the formalism of effective field theory, in which one allows for the addition of arbitrary terms to the Lagrangian, provided that they are consistent with the desired symmetries, such as general covariance.
How the covariant cutoffs considered here and effective field-theoretic treatments of inflation \cite{Cheung:2007st} compare is discussed in \cite{Chatwin-Davies:2022irs}.}
In particular, we could soften the cutoff by choosing a function $f$ that smooths out the Heaviside step function.
Even so, the cutoff remains covariant, since it is obtained by restricting the spectrum of the scalar operator $\Box$.
For now we will continue to focus on a sharp cutoff at $\Omega$ and we will return to soft cutoffs later.

\section{Corrections to primordial power spectra}

Now let us assume that $\mathcal{M}$ is a spatially flat Friedmann-Lema{\^i}tre-Robertson-Walker (FLRW) spacetime with the line element
\begin{equation}
    \dee s^2 = a^2(\eta)\left(-\dee \eta^2 + \dee {\bm x}^2 \right),
\end{equation}
where $a(\eta)$ is a scale factor describing slow-roll inflation, i.e. the spacetime is quasi-de Sitter. Our goal is to compute the correction, $\delta \Delta_\mathcal{R}^2(k)$, to the PPS due to a sharp covariant cutoff $\Omega$ imposed on the spectrum of the FLRW d'Alembertian $\Box$.
Notice that a spatial Fourier transform with respect to $\bm x$ preserves the spectrum of $\Box$. Explicitly, if $\Box u(\eta, {\bm x}) = \lambda u(\eta, {\bm x})$, then $\Box_{\bm k} u(\eta, {\bm k}) = \lambda u(\eta, {\bm k})$, where $u(\eta,\bm k)$ is the spatial Fourier transform of $u(\eta,\bm x)$ and $\Box_{\bm k} = -a^{-4}\partial_\eta(a^2 \partial_\eta)-k^2 a^{-2}$ is the spatial Fourier transform of $\Box$.
We may therefore impose the cutoff on each mode $k = |{\bm k}|$ individually by cutting off the spectrum of each $\Box_{k}$.

Let us now compute what the effect of such a covariant cutoff is on the primordial power spectrum $\DR(k)$. From \Eq{eq:PSP-GF}, one finds that the relative correction is
\begin{equation}
    \label{eq:reldiffscalar}
    \frac{\delta \Delta_\mathcal{R}^2(k)}{\Delta_\mathcal{R}^2(k)} = \mrm{Re} \left( \frac{\delta G_F(\eta_k,k)}{G_F(\eta_k,k)} \right) + O(\delta G_F^2),
\end{equation}
where $\delta G_F = G_F^\Omega - G_F$. Thus, to compute the relative correction to the PPS at leading order in $\delta G_F$, we need to compute $G_F$ and $\delta G_F$ for a quasi-de Sitter spacetime.

Let us first consider $G_F$.
For a general scale factor, a closed-form expression for $G_F$ is not known, although it can be expressed as a mode sum.
Computing the Feynman propagator $G_F^\mrm{dS}$ for a free scalar field on a pure de Sitter background is a standard computation, however, and the answer can be obtained in closed form \cite{Birrell:1982ix}. 
In a quasi-de Sitter spacetime, one can thus employ a \textit{slow-roll approximation} for $G_F$, which approximates the exact slow-roll calculation with a corresponding de Sitter calculation.
More precisely, we approximate
\begin{align}
    G_F(\eta_k,k)\approx G_F^\mrm{dS}(\eta_k^\mrm{dS},k),
\end{align}
where $\eta^\mrm{dS}_k = -1/k$ is the de Sitter horizon crossing time for a mode $k$ and where, in computing $G_F^\mrm{dS}$, we take the de Sitter Hubble constant $H^\mrm{dS}$ to be equal to $H(\eta_k)$, the instantaneous Hubble parameter in the quasi-de Sitter spacetime at the moment when the mode $k$ crosses the horizon. By equating Eqs.~\eqref{eq:scalar-spectrum-pheno} and \eqref{eq:scalar-spectrum-slow-roll} and expanding $\epsilon$ to leading order about the pivot scale, we obtain an expression for $H(\eta_k) \equiv H(k)$ in terms of observable parameters,
\begin{equation} \label{eq:Heff}
    H(k) = \Mpl\sqrt{\pi A_s \epsilon_\star} \left( \frac{k}{k_\star}  \right)^{-\epsilon_\star},
\end{equation}
where CMB measurements \cite{Planck:2018jri} constrain the value of $\epsilon$ at the pivot scale to $\epsilon_\star \lesssim 0.004$.
The relative error incurred through the use of the slow-roll approximation is of the same order as the slow-roll parameters \cite{Chatwin-Davies:2022irs}, so the approximation is accurate in the regime where these parameters are much smaller than 1.

Next let us discuss $\delta G_F$. Defining the projector $P_\Omega^\perp \equiv I - P_\Omega$, which projects onto the eigenspace of $\Box_k$ with eigenvalues $|\lambda|>\Omega^2$, we can write 
\begin{align}
    \delta G_F = P_\Omega^\perp G_F P_\Omega^\perp - P_\Omega^\perp G_F - G_F P_\Omega^\perp.
\end{align}
Computing $\delta G_F$ is conceptually straightforward:
For each mode $k$, one solves the the Sturm-Liouville eigenvalue problem $\Box_k \psi_\lambda = \lambda \psi_\lambda$ to obtain an orthonormal eigenfunction basis.
One then constructs the projectors $P_\Omega^\perp$ and uses them to compute $\delta G_F$.
Once again, however, this calculation is intractable for general slow-roll spacetimes.
Fortunately, computing the de Sitter correction $\delta G_F^\mrm{dS}$ is more tractable, and hence we can accurately approximate $\delta G_F$ for a quasi-de Sitter spacetime by applying the slow-roll approximation here as well \cite{Chatwin-Davies:2022irs}.

A technical complication which arises in computing $\delta G_F^\mrm{dS}$ is that the minimal realization of $\Box_k$ as a differential operator is only symmetric and not self-adjoint.
In functional analytic language, its deficiency indicies are $(1,1)$ and a generalized boundary condition must be specified \cite{naimark1968linear}.
This freedom in defining a self-adjoint $\Box_k$ is in correspondence with the choice of vacuum state for de Sitter.
We fix this freedom by requiring that $\Box_k$ act as the left inverse of $G_F$, where the latter is determined from canonical quantization and from having chosen the Bunch-Davies state as the vacuum; see \cite{Chatwin-Davies:2016byj,Chatwin-Davies:2022irs} for further details.
In this way we are able to compute $\delta G_F^\mrm{dS}$, and hence, via the slow-roll approximation, $\delta G_F$ for a quasi-de Sitter spacetime.
We chose the Bunch-Davies state so as to minimally diverge from the standard inflationary calculation and to avoid possible issues with gauge invariance and perturbative divergences \cite{Urakawa:2010kr}.
Other choices are possible, but the resulting predictions may differ significantly from those presented here \cite{deBlas:2016puz,Navascues:2021qcp}.

Upon computing $G_F$ and $\delta G_F$, we can use Eq. \eqref{eq:reldiffscalar} to construct the relative correction $\dDR/\DR$ to the primordial power spectrum due to a sharp covariant cutoff at the scale $\Omega$. We find that $\delta \Delta_\calR^2/\Delta_\calR^2$ is a function of only $\sigma(k) \equiv H(k)/\Omega$, the ratio of the Hubble scale \eqref{eq:Heff} and cutoff scale.\footnote{Strictly speaking, $\calR$ experiences a modified Hubble parameter due to a modified scale factor $z = (\Mpl^2\epsilon/4\pi)^{1/2}a$; however, $H = \dot{a}/a$ and $\dot{z}/z$ coincide to leading order with deviations that are suppressed by slow-roll parameters.}
Concretely, the correction to the PPS is
\begin{equation} \label{eq:prediction}
   \frac{\delta \Delta^2_\calR}{\Delta^2_\calR}=
    \mathcal{C}
    \frac{\sigma(k)^{3/2}}{\ln(\sigma(k)/2)} \sin\left(\omega(k)\,  \sigma(k)\right),
\end{equation} 
where\footnote{More precisely, $\mathcal C = 2(\cos1+\sin1)\pi^{-1}$, where $\cos1$ and $\sin1$ appear from evaluating $J_{3/2}(k\eta)$ and $Y_{3/2}(k\eta)$ at horizon crossing, $k\eta=-1$.} $\mathcal{C} = 0.8796...$ and where we have defined
\begin{equation} \label{eq:frequency}
    \omega(k) \equiv \frac{1}{\sigma(k)^2}\ln\frac{e\sigma(k)}{2}.
\end{equation}

\begin{figure}[ht]
    \centering
    \includegraphics[width=0.42\textwidth]{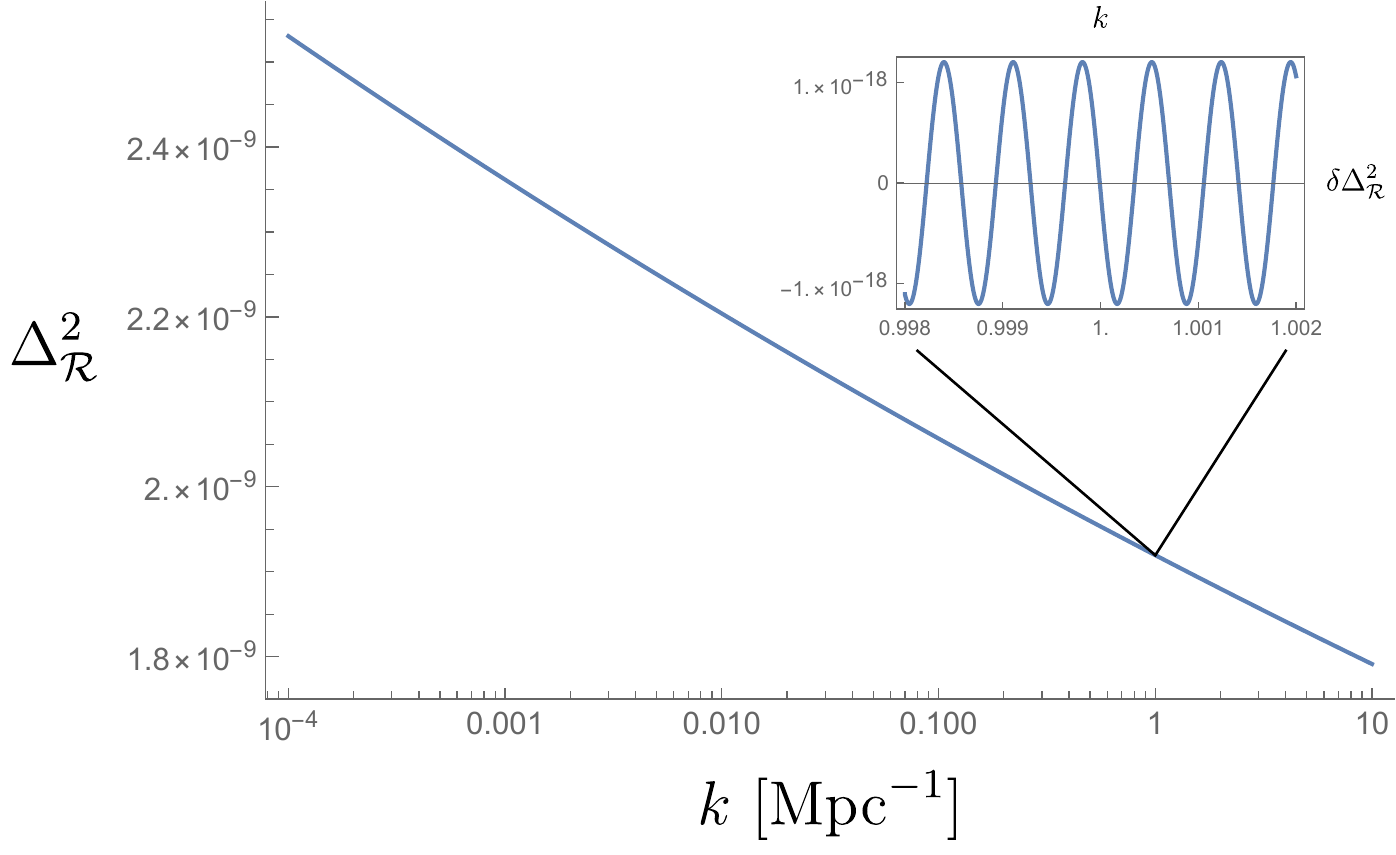} \\
    \includegraphics[width=0.42\textwidth]{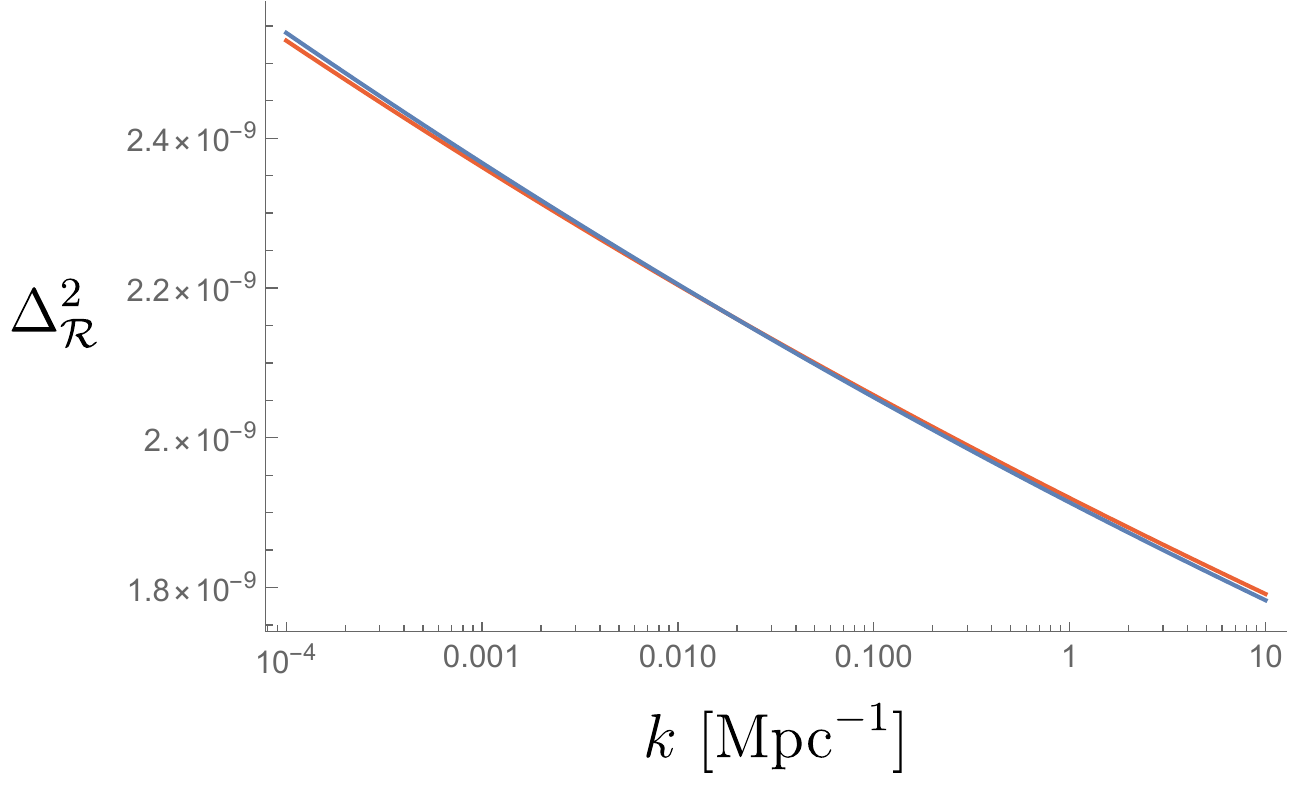}
    \caption{The primordial scalar power spectrum, corrected due to a sharp cutoff at $\Mpl / \Omega = 1$ (upper plot) and at $\Mpl / \Omega = 2.1 \times 10^4 $ (lower plot). The inset in the upper plot shows the correction itself, $\delta \Delta_\calR^2$, over a small range of $k$ such that its oscillations are resolved.
    The uncorrected power spectrum is also shown in red in the lower plot.
    The parameter values used are $A_s = 2.1 \times 10^{-9}$, $k_\star = 0.05~\mrm{Mpc}^{-1}$, $n_s = 0.97$, and $\epsilon_\star = 0.003$.}
    \label{fig:prediction}
\end{figure}

The correction $\dDR(k)/\DR(k)$ to the PPS thus consists of oscillations whose amplitude and frequency are functions of $k$; see Fig. \ref{fig:prediction}. 
Notice that, besides the choice of pivot scale $k_\star$ and the tightly-constrained parameter $A_s$, the correction depends only on the first slow-roll parameter $\epsilon_\star$, for which only an upper bound is known, and $\Omega$, the energy scale at which we impose a sharp, covariant UV cutoff.
The amplitude of the oscillations scales like $\Omega^{-3/2}\epsilon_\star^{3/4}$ (up to a logarithmic factor) and the frequency in the $k$ domain scales like $\Omega \epsilon_\star^{1/2}$.
If the cutoff is brought very close to the Hubble scale and for values of $\epsilon_\star$ that are not too small, the correction becomes so prominent that existing observational data should be able to place simultaneous bounds on the value of the first slow-roll parameter and where the cutoff scale could lie.

The analysis so far has been focused on a sharp UV cutoff.
As discussed earlier, by using a continuous function $f(\lambda)$ instead of $\theta$ in Eq. \eqref{eq:cutoff}, it is also possible to soften the cutoff. Specifying such cutoffs necessarily requires more than just a single new parameter $\Omega$, and so the corresponding class of corrections $\dDR/\DR$ to the PPS will depend on multiple parameters. Our analysis in \cite{Chatwin-Davies:2022irs} shows that the effect of smoothing out the cutoff is to alter the amplitude and phase of the corrections to the PPS in a way that depends on the specific form of the smoothing. In particular, the amplitude of the correction tends to decrease as one increases the smoothness of the cutoff.
However, we also found that if there exists a natural covariant UV cutoff at a well-defined finite scale, then the frequency of the predicted oscillations in the PPS, given by Eq. \eqref{eq:frequency}, is essentially independent of whether the cutoff is sharp or softened.

\section{Discussion}

If, as is widely assumed, there exists a natural UV cutoff at or close to the Planck scale, then the framework of quantum field theory should increasingly exhibit the presence of this UV cutoff when approaching the cutoff scale from low energies. Since the quantum field theoretic calculations of cosmic inflation involve  scales that are only about 5-6 orders of magnitude from the Planck scale, the inflationary predictions might exhibit a noticeable  impact of such a natural UV cutoff. 

Here, we presented the first explicit inflationary predictions for a natural UV cutoff within the inflationary path integral that is covariant. We focused on the case of the scalar primordial power spectrum, $\DR(k)$, and we considered both the simplest case where the UV cutoff is sharp, and also the case where the cutoff is a softened decay. 
The full details of the calculations, as well as the corresponding predictions for the tensor power spectrum, are discussed in \cite{Chatwin-Davies:2022irs}.

Our predictions assume single-field inflation but are otherwise model independent. We do not make assumptions on the inflationary potential; instead, we write our predictions explicitly in terms of measured cosmological parameters.

We found that $\DR(k)$ is corrected by small oscillations, $\dDR(k)$, on top of the power spectrum. 
Logarithmic oscillations in and of themselves are not unique to our prediction---see, e.g., Refs.~\cite{Easther:2005yr,Ashtekar:2021izi,Agullo:2021oqk}, as well as Ref.~\cite{Calcagni:2016ofu} and Refs.~[27-43] therein---however, the highly specific signature \eqref{eq:prediction} and the chirping frequency \eqref{eq:frequency} are, to our knowledge, unique.
Propagated forward in time, they would imply corresponding corrections to the CMB and to large-scale cosmic structure.
We remark that a long wavelength oscillation that amounts to a trough at large comoving scales would reduce the power in the PPS there, which is in principle compatible with an apparent lack of power at large scales in measured CMB power spectra \cite{Perivolaropoulos:2021jda}.

The most robust prediction here is for the predicted frequency of the oscillations, \Eq{eq:frequency}. This is because this frequency is virtually independent of the exact form of the covariant cutoff, i.e., whether it is a sharp cutoff or whether the cutoff turns on smoothly. Instead, the predicted frequency depends only on the cutoff via its characteristic scale $\Omega$.
If the UV cutoff is sharp, then the amplitude and the phase of the oscillations also are fixed entirely by $\Omega$ and the cosmological parameters $A_s$ and $\epsilon_\star$.
In this case, we obtained a simple analytic expression for $\dDR(k)$, given in Eq. \eqref{eq:prediction}.

The specificity of our predictions might help significantly in testing them, in particular by using template search methods. 
This would be similar to recent measurements of gravitational radiation emitted by distant massive objects \cite{LIGOScientific:2016aoc}, which succeeded by searching for members of a three-parameter family of template waveforms \cite{Privitera:2013xza}.

If the natural UV cutoff lies too close to the Planck scale, then the resultant fast oscillations risk being washed out in the processes of binning and computing derived power spectra, such as the primordial power spectrum, from measured CMB power spectra.
Conversely, the corrections that the UV cutoff induces are significantly easier to detect the closer $\Omega$ is to the Hubble scale rather than to the Planck scale.
This is because, towards the Hubble scale, the amplitude of the oscillations increases while their frequency decreases.

It will be very interesting to determine which values for a natural covariant UV cutoff $\Omega$ can already be ruled out, namely by comparing the predictions presented here with the experimentally inferred primordial power spectrum.
In the absence of an independent measurement of the first slow-roll parameter, such bounds will necessarily jointly depend on $\epsilon_\star$, since the latter is itself only upper-bounded by existing data and its value also affects the oscillations' amplitude and frequency.
As future experiments increase the accuracy with which the primordial power spectrum is known, a positive detection of the signature of a natural UV cutoff may even come within the reach of observations.

\begin{acknowledgments}
We thank Panos Betzios, Fran\c cois Bouchet, Richard Easther, Simon Foreman, Lukas Hergt, Arjun Kar, Jorma Louko, Rob Martin, and Mark Van Raamsdonk for helpful discussions during the preparation of this manuscript, as well as Gianluca Calcagni and Albert Roura for comments on the first version.
ACD acknowledges the support of the Natural Sciences and Engineering Research Council of Canada (NSERC), [funding reference number PDF-545750-2020].
ACD was supported for a portion of this work as a postdoctoral fellow (Fundamental Research) of the National Research Foundation -- Flanders (FWO), Belgium. AK acknowledges support through a Discovery Grant of the National Science and Engineering Council of Canada (NSERC) and a Discovery Project grant of the Australian Research Council (ARC). PS acknowledges support from the NSERC CGS-D award.

~
\end{acknowledgments}

\bibliographystyle{utphys-modified}
\bibliography{refs.bib}

\end{document}